\documentclass[aps,prd,twocolumn,superscriptaddress,groupedaddress]{revtex4}
\usepackage{graphicx}  
\usepackage{dcolumn}   
\usepackage{bm}        
\usepackage{amssymb}   
\usepackage{amsmath}
\usepackage{xcolor}
\usepackage{rotating}
\usepackage{amsthm}
\usepackage{subfig}
\usepackage{subfloat}
\usepackage{actuarialsymbol}
\usepackage{braket}
\usepackage{mathtools}
\usepackage{bbold}
\usepackage{amssymb}
\usepackage{enumitem}
\usepackage{comment}
\usepackage{appendix}

\newcommand{\be}{\begin{equation}}\newcommand{\ee}{\end{equation}}
\newcommand{\bea}{\begin{eqnarray}}\newcommand{\eea}{\end{eqnarray}}
\newcommand{\brr}{\begin{array}}\newcommand{\err}{\end{array}}
\newcommand{\bit}{\begin{itemize}}\newcommand{\eit}{\end{itemize}}
\newcommand{\ben}{\begin{enumerate}}\newcommand{\een}{\end{enumerate}}

\newcommand{\ba}{\begin{array}}
\newcommand{\ea}{\end{array}}

\definecolor{darkred}{rgb}{.8,0,0}

\definecolor{darkblue}{rgb}{0,0,.7}

\def\Ga{\Gamma}

\def\Om{\Omega}

\def\1{{_{1}}}\def\2{{_{2}}}

\def\noHe0{:\;\!\!\;\!\!:H_e(0):\;\!\!\;\!\!:}
\def\noHm0{:\;\!\!\;\!\!:H_\mu(0):\;\!\!\;\!\!:}

\def\Ga{\Gamma}

\def\Om{\Omega}

\def\1{{_{1}}}\def\2{{_{2}}}

\usepackage{ragged2e}

\begin{document}
\title{On the possibility of superradiant neutrino  emission by atomic condensates}

\author{Massimo Blasone}
\email{blasone@sa.infn.it}
\affiliation{Dipartimento di Fisica, Universit\`a di Salerno, Via Giovanni Paolo II, 132 I-84084 Fisciano (SA), Italy}
\affiliation{INFN, Sezione di Napoli, Gruppo collegato di Salerno, Italy}

\author{Loredana Gastaldo}
\email{Gastaldo@kip.uni-heidelberg.de}
\affiliation{Kirchhoff Institute for Physics, Heidelberg University, Germany}

\author{Francesco Romeo}
\email{fromeo@unisa.it}
\affiliation{Dipartimento di Fisica, Universit\`a di Salerno, Via Giovanni Paolo II, 132 I-84084 Fisciano (SA), Italy}
\affiliation{INFN, Sezione di Napoli, Gruppo collegato di Salerno, Italy}

\begin{abstract}

 In a recent work [B. J. P. Jones and J. A. Formaggio, Phys. Rev. Lett. 135, 111801 (2025)], the possibility of superradiant neutrino emission from atomic condensates has been theoretically proposed. Subsequent analysis by Y. K. Lu, H. Lin, and W. Ketterle [arXiv:2510.21705] questioned this scenario, emphasizing the limiting role of the fermionic nature of the decayed atoms. In this study, we revisit the problem and discuss under which conditions collective emission phenomena might still emerge in cold-atom systems.
\end{abstract}

\maketitle

\section{Introduction}

The prospect of generating coherent radiation outside the electromagnetic sector represents an open
direction in quantum many-body physics. In particular, the possibility of realizing collective or even
stimulated neutrino emission has attracted recent interest \cite{Jones:2024lhf,Lu:2025sbp}, as it would open a qualitatively new regime
of weak-interaction physics while offering a unique window into quantum coherence in neutrino
production processes.

Jones and Formaggio~\cite{Jones:2024lhf} argued that a Bose–Einstein condensate (BEC)
of radioactive atoms could, under suitable conditions, exhibit a form of neutrino superradiance
analogous to Dicke-enhanced photon emission \cite{dicke}. Their analysis highlighted a potential amplification
mechanism but relied on analogy-based arguments and on an effective single-channel treatment of the
decay. Shortly thereafter, a critical assessment by Lu, Lin and Ketterle~\cite{Lu:2025sbp} emphasized that,
for realistic electron-capture isotopes, the fermionic statistics of the daughter atoms leads to severe
Pauli blocking, which would suppress the cooperative enhancement presumed in the original scenario.
These observations raise the broader question of whether superradiant neutrino emission may survive
in alternative settings, or whether the underlying idea is fundamentally incompatible with the
statistics of available decay channels.

The aim of this work is to clarify this point by developing and analyzing a set of minimal, yet
microscopically motivated, models that isolate the statistical constraints governing cooperative
emission. Instead of focusing on a single decay channel, we consider  three statistically distinct possibilities: bosonic atoms decaying into bosonic species which could appear for gamma decay of metastable nuclear excited states, bosonic atoms decaying into fermions, and fermionic atoms (in the deep BEC regime) decaying into bosons, these last two options are accessible via electron capture. Each case captures a different
statistical pathway and allows us to quantify whether and how bosonic stimulation or Pauli blocking
affects the onset of collective emission. Our approach, based on Lindblad dynamics and effective
few-mode Hamiltonians derived from a real-space microscopic model, provides a controlled framework
within which the competing roles of cooperativity, statistics, and dissipation can be assessed.

We find that while the boson-to-fermion channel is indeed strongly suppressed by Pauli blocking,
consistent with Ref.~\cite{Lu:2025sbp}, cooperative behaviour can persist in scenarios beyond those
considered so far—most notably when fermionic atoms in the deep BEC regime decay into bosonic
species. This indicates that the conceptual idea underlying Ref.~\cite{Jones:2024lhf}, which is realized by a boson-to-boson model, is not ruled out on fundamental grounds, but that its realization requires physical systems with suitable decay pathways.

The paper is organized as follows. In Section~\ref{sec:EC} we review the essential features of the
electron-capture process, with emphasis on how it constrains the statistics of the decay products.
Section~\ref{sec:BosBos} introduces a minimal model in which bosonic atoms in a BEC decay into a
bosonic daughter species, reproducing the superradiant neutrino-emission mechanism  of
Ref.~\cite{Jones:2024lhf}. In Section~\ref{sec:BosFerm} we examine the complementary case in which
bosonic atoms decay into fermionic products; here we recover the central conclusion of
Ref.~\cite{Lu:2025sbp}, namely the suppression of cooperative emission due to Pauli blocking.
Section~\ref{sec:FermBos} then considers a scenario where fermionic atoms in the deep BEC regime
decay into bosonic species, a situation in which a pronounced superradiant behaviour emerges.
Finally, Section~\ref{sec:Concl} summarizes our findings and discusses prospects for the possible
experimental realization of the scenarios analyzed in this work.

\section{Electron Capture}
\label{sec:EC}
Before discussing about the possibility to have enhanced decay rate for electron capture in BEC of suitable radioactive isotopes, it is important to spend a few words on the electron capture process. Nuclei with high proton number could undergo a pure electron capture process if the energy available for the decay, given by the difference between the mass of the parent atom and the mass of the daughter atom, is smaller than 1022 keV, the threshold for the competing $\beta ^+$-decay. If this energy constraint is fulfilled, then an electron of the inner shells can be captured in the nucleus where a proton turns to a neutron and an electron neutrino is emitted. The vacancy in the electron shells corresponds to a particular excitation energy of the daughter atom, which is then released within fs. 

The first important aspect to notice is that the spin of the electron system for parent and daughter atoms will change from integer to half-integer or viceversa while the nucleus will maintain the spin either integer or half-integer given the fact that the number of nucleons does not change. This aspect imposes limits on the stability of the condensate after electron capture. 

A second important fact to be considered is the energy spectrum of neutrinos. The total energy available to the decay is shared among the neutrino, the atomic excitation energy and the nuclear recoil. The atomic excitation energy is, in first approximation, given by the binding energy of the captured electron in the field generated by the daughter nucleus. The capture probability is proportional to the probability to find the electron in the nucleus, therefore it is proportional to square of the electron wavefunction calculated at the nucleus. Electron from the $s$ and $p_{1/2}$ shells have the highest probability to be captured. The important implication of this is that the neutrino energy spectrum is not monochromatic, but is characterized by several resonances with intrinsic width in the eV scale and not negligible tails. Nevertheless, for the majority of the decays the capture of $1s$ electrons is dominant and can reach a fraction greater than $80\%$. 

Two prominent examples of electron capture spectra which have been recently studied with high precision are the one of $^{163}$Ho \cite{ECHo1, ECHo2, HOLMES1, HOLMES2} and $^7$Be \cite{BEEST1, BEEST2}. In the first case, the parent atom is a boson while the daughter atom $^{163}$Dy is a fermion. The energy available for the decay is just $Q = 2863.2(6)$ eV \cite{PTMS2024} which allows capture only from the $3s$-shell, while the halflife is about 4570 years \cite{Baisden}. A theoretical description of the electron capture in $^{163}$Ho using an $ab-initio$ approach was presented in \cite{Brass1, Brass2}. Although the intensity of the tails is still not perfect, the discussed model well captures the several structures of the spectrum and, in particular, the multiplet nature of the main resonances. The nuclear recoil of the daughter atom $^{163}$Dy,  due to the size of the nucleus and the small energy of the emitted neutrinos, is of the order of a few $\mu$eV. 
The $^7$Be has, on the other hand, very different properties. First of all, the parent atom is a fermion and, consequently, the daughter atom is a boson. The halflife of this decay is about 52 days and could be suitable for testing decay rate enhancement. At the same time, the energy available to the decay is $Q = 861.963(23)$ keV \cite{Bhandari} which leads to a remarkable recoil energy for the daughter nucleus $^7$Li of about 50 eV affecting the stability of the condensate. While in the electron capture of $^7$Be more than 90$\%$ of the decay occurs via capture of $1s$ electrons, it is also important to mention that a small fraction of decays, of the order of $10\%$, populates a short lived excited nuclear state in $^7$Li with halflife $\tau_{1/2}$ = 728(20) fs which relaxes to the ground state via the emission of a $\gamma$-ray with energy 477.603(2) keV.

An analysis of the nuclide chart for electron captures in which the parent atom decays to a stable atom in the ground state with halflife longer than one day and $Q$-value smaller than 1022 keV showed that the candidates are typically nuclei with an odd number of nucleons, therefore with semi-integer nuclear spin. The number of nuclides whose decays belong either to boson to fermion or fermion to boson group is equivalent and cover a wide range of decay-rates and decay energies.

The microscopic picture outlined above illustrates that electron capture is a multi–scale process involving changes in atomic statistics, a structured neutrino spectrum, and fast electronic and possibly also nuclear relaxation channels. A complete dynamical treatment of all these ingredients is beyond present analytical control and would require a fully ab initio description of the coupled atomic, nuclear, and neutrino degrees of freedom. For this reason, in the remainder of this work we resort to minimal effective models that retain only the degrees of freedom most relevant for assessing whether cooperative decay can emerge, neglecting the energy scale of the nuclear recoil and the energy released in the atomic de-excitation process.
Among the essential features that must be preserved are the quantum statistics of both the parent and
daughter atoms, and the manner in which these statistics influence bosonic stimulation or Pauli blocking.
In the following sections, we formulate controlled simplified models tailored to isolate the statistical constraints imposed by the decay process and to quantify their impact on the possibility of realizing a superradiant neutrino–emission regime in ultracold atomic systems.

\section{Neutrino superradiance in a bosonic-bosonic model}
\label{sec:BosBos}

In this section, we introduce an effective microscopic model aimed at 
capturing the seminal idea, originally put forward by Jones and Formaggio,
that a radioactive atomic Bose--Einstein condensate (BEC) could serve as a 
source of neutrinos exhibiting a form of superradiant amplification.  
It is important to emphasize, however, that the analogy with photon 
superradiance is necessarily incomplete.  
Neutrinos obey fermionic statistics and cannot be trapped or confined in a 
cavity; as a consequence, phase coherence in the emitted field cannot develop 
in the standard Dicke sense, and the phenomenon effectively reduces to a 
collective avalanche emission, whenever such a regime is dynamically 
accessible.  This limitation, already noted in Ref.~\cite{Jones:2024lhf},
must be kept in mind when discussing the possible cooperative enhancement of 
the decay rate.

Starting from a real-space microscopic description of the emitting region, i.e. the 
spatial domain hosting the radioactive condensate, one can derive an 
effective low-energy Hamiltonian expressed in terms of three collective modes:  
a bosonic operator $a$ describing the initial condensate, a bosonic operator 
$b$ associated with the daughter atoms produced via electron capture, and a 
fermionic operator $c$ representing the neutrino field.
A complete derivation of this reduction, including the approximations leading 
from the full many-body Hamiltonian to the single-mode description of the neutrino field adopted 
below, is detailed in Appendix~A.

Within this effective framework, the coherent part of the dynamics is governed 
by the Hamiltonian
\begin{equation}
H = \epsilon_a\, a^\dagger a 
  + \epsilon_b\, b^\dagger b 
  + E_\nu\, c^\dagger c 
  + g\, c^\dagger b^\dagger a 
  + g^\ast a^\dagger b\, c ,
\label{eq:H_eff}
\end{equation}
where $\epsilon_a$ and $\epsilon_b$ denote the effective energies of the 
initial and final atomic modes, and $E_\nu$ is the neutrino energy.  
The effective vertex term encodes the conversion of one bosonic atom in the 
condensate into a daughter boson and an emitted neutrino.
In order to consider a fermionic leaking channel corresponding to the neutrino emission, we adopt a Lindblad formalism according to which the time-evolution of a generic operator $O$ in Heisenberg picture takes the form \cite{BreuerPetruccione2002}:
\begin{eqnarray}
\frac{dO}{dt}=\frac{i}{\hbar}[H,O]+ \sum_n \Big(L_n^{\dagger} O L_n-\frac{1}{2}\{L^{\dagger}_nL_n,O\} \Big),
\end{eqnarray}
with $L_n$ a set of jump operators.

Taking into account a single dissipation channel with dissipator $L=\sqrt{\gamma} \,c$ and dissipation rate $\gamma=2\Gamma/\hbar$, the time evolution of the atomic populations is described by the following equations:
\begin{eqnarray}
    \frac{d}{dt} \braket{n_a} &=&
    \frac{i}{\hbar} \left[ g \braket{c^\dagger   b^\dagger a} 
     - g^*\braket{a^\dagger  b c }\right],
  \\
  \frac{d}{dt} \braket{n_b} &=&
    \frac{i}{\hbar} \left[ -g \braket{c^\dagger   b^\dagger a} 
     + g^*\braket{a^\dagger  b c }\right],
\end{eqnarray}
from which one immediately obtains the conservation law $d/dt (\braket{n_a}+\braket{n_b})=0$. 

The dynamics of the populations depends on the correlation function $S\equiv \braket{a^\dagger  b c } $, with $S^*=\braket{c^\dagger   b^\dagger a} $. Thus the equation for $n_a$ can be written as:
\begin{equation}
    \braket{\dot{n}_a} =\frac{i}{\hbar} \left[ g S^*
     - g^*S\right]  = \frac{2}{\hbar} \Re\left[i g S^*\right] =-\frac{2}{\hbar} \Im\left[g S^*\right],
\end{equation}
which, taking into account the conservation of the total number of atoms, also implies:
\begin{equation}
    \braket{\dot{n}_a} =-\braket{\dot{n}_b} =\frac{2}{\hbar} \Im\left[g^* S\right].
\end{equation}
The average number of neutrinos in the volume of the emission region is obtained by considering the Lindblad-Heisenberg equation of motion for $n_c=c^\dag c$. Thus, the time evolution of $\braket{n_c}$ is regulated by the following equation:
\begin{equation}
    \braket{\dot{n}_c} =-\frac{2}{\hbar} \Im\left[g^* S\right]
    -\frac{2 \Ga}{\hbar} \braket{n_c},
\end{equation}
where the second term in the rhs of the equation accounts for the neutrino escaping from the emission region.\\
Here $\Ga \approx \hbar c/\ell$, with $c$ the speed of light and $\ell$ the typical length of the condensate cloud, can be identified with the largest energy scale of the system, so that the instantaneous mean number of neutrinos which are present in the emission region is well approximated by the quantity $\braket{n_c}^{eq}\approx - \Im\left[g^* S\right] /\Ga$.

The populations dynamics is evidently affected by $S(t)=\braket{a^\dag b c}$, whose equation of motion is given by:
\begin{eqnarray}\nonumber
    \dot{S} &=& \frac{i}{\hbar} \Big[ (\Delta + i \Gamma)S  
\\ 
&&  + g \braket{n_c n_b(1+n_a) - n_a(1+n_b)(1-n_c)}\Big],
\end{eqnarray}
with $\Delta=\epsilon_a-\epsilon_b-E_{\nu}$.\\ 
Hereafter, we neglect high order correlations and adopt the following factorization:
\begin{eqnarray}
    \braket{n_c n_b(1+n_a)} &\approx &\braket{n_c} \braket{n_b} (1+\braket{n_a})
    \\ [2mm]
    \braket{n_a(1+n_b)(1-n_c)}&\approx &
    \braket{n_a} (1+\braket{n_b})(1-\braket{n_c}), \quad 
\end{eqnarray}
so that we obtain:
\begin{eqnarray} \nonumber
     \dot{S} &=&\frac{(i\Delta - \Gamma)}{\hbar}  S
    + \frac{i g}{\hbar}\braket{n_c} \braket{n_b} (1+\braket{n_a})
    \\ 
    && - \frac{i g}{\hbar}\braket{n_a}(1+\braket{n_b})(1-\braket{n_c}),
\end{eqnarray}
which is clearly affected by the Pauli blocking term $(1-\braket{n_c})$.

An important simplification of the equation of motion of $S$ is obtained observing that $\braket{n_c}\ll 1$, in view of the fact that $\Gamma\gg 1$. Following the above line of reasoning, we obtain the approximate equation:
\begin{equation}
\label{eq:SEOM}
    \dot{S} =\frac{i}{\hbar} \left[ (\Delta + i \Gamma)S
    - g \braket{n_a} (1+\braket{n_b})\right].
\end{equation}
Assuming that $\Gamma\gg 1$, the solution of Eq.~\eqref{eq:SEOM} can be accurately approximated by the expression:
\begin{equation}
    S \approx \frac{i g\braket{n_a} (1+\braket{n_b}) }{i\Delta -\Gamma}.
\end{equation}

\begin{figure}[t]
    \includegraphics[width = 8.5 cm]{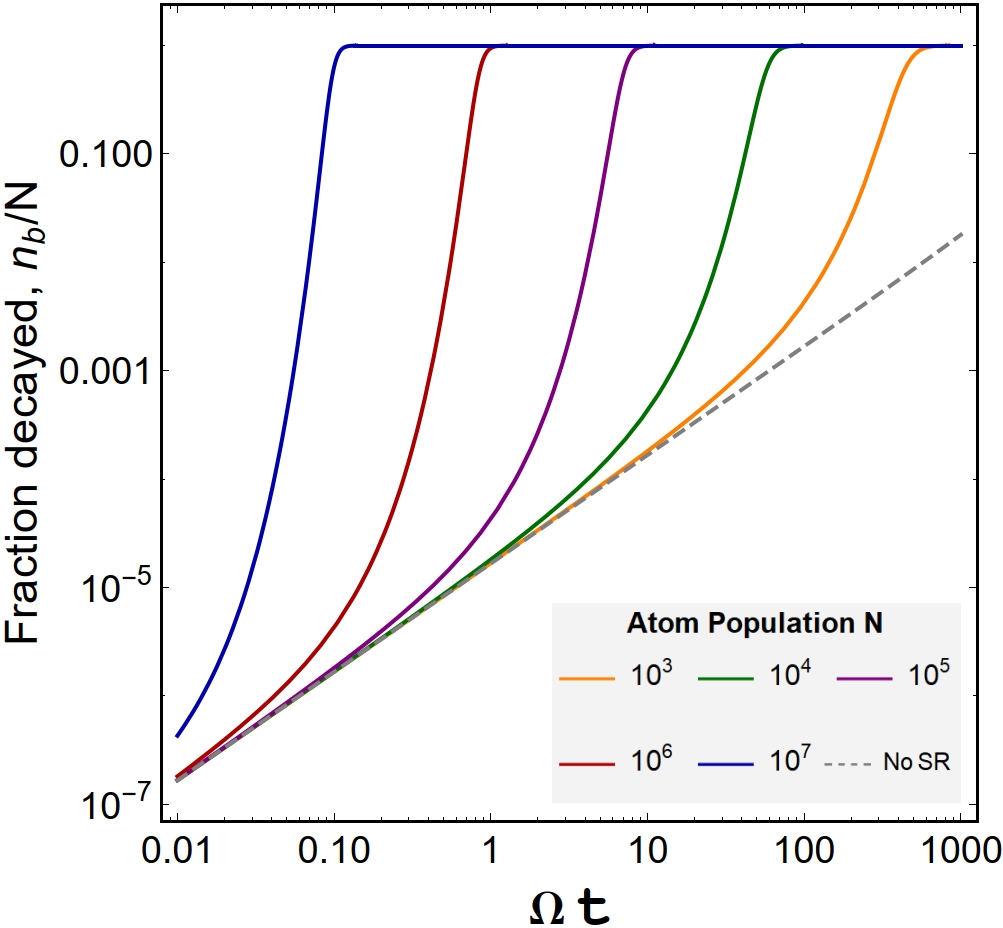}
    \caption{Plot of the decayed fraction $\frac{\braket{n_b}}{N}$, according to Eq.\eqref{nbsolution}, for different atom populations $N$ as specified in figure legend (See Ref.\cite{Jones:2024lhf} for comparison). Dashed line corresponds to the case of no superradiance.}
    \label{fig1}
\end{figure}

Using the above relation, the dynamics for $\braket{n_b}$ takes the following form:
\begin{eqnarray}
    \braket{\dot{n}_b} &=&
        -\frac{2}{\hbar} \Im\left[\frac{|g|^2 \braket{n_a} (1+\braket{n_b})}{\Delta +i\Gamma}\right].
   \end{eqnarray}
The latter, can be rewritten in the form:
\begin{eqnarray}\nonumber
\label{eq:nbDyn}
    \braket{\dot{n}_b} &=& 
     \frac{2 |g|^2}{\hbar} \frac{\Gamma }{\Delta^2 +\Gamma^2} \braket{n_a} (1+\braket{n_b})
     \\ [2mm]
     &=&      \Om \, ( 1+\braket{n_b})\, (N-\braket{n_b}),
\end{eqnarray}
where we used $\braket{n_a}+\braket{n_b}=N$ and set $\Om\equiv \frac{2 |g|^2}{\hbar} \frac{\Gamma }{\Delta^2 +\Gamma^2}\approx \frac{2|g|^2}{\hbar \Gamma}$. Such an equation exhibits logistic growth of the number $\braket{n_b}$, with solution
\begin{equation}
\label{nbsolution}
    \frac{\braket{n_b}}{N}\,=\, \frac{e^{\Om (N+1)t}-1}{e^{\Om (N+1)t}+N}\,,
\end{equation}
derived with initial condition $\braket{n_b}_{t=0}=0$.\\
Equation \eqref{nbsolution}, whose behavior is exemplified in Fig. \ref{fig1}, reproduces exactly the superradiant behavior discussed in Eq. (27) of Ref. \cite{Jones:2024lhf}.

\subsection{Effects of atomic interaction}
 
\begin{figure}[t]
    \includegraphics[width = 8.5 cm]{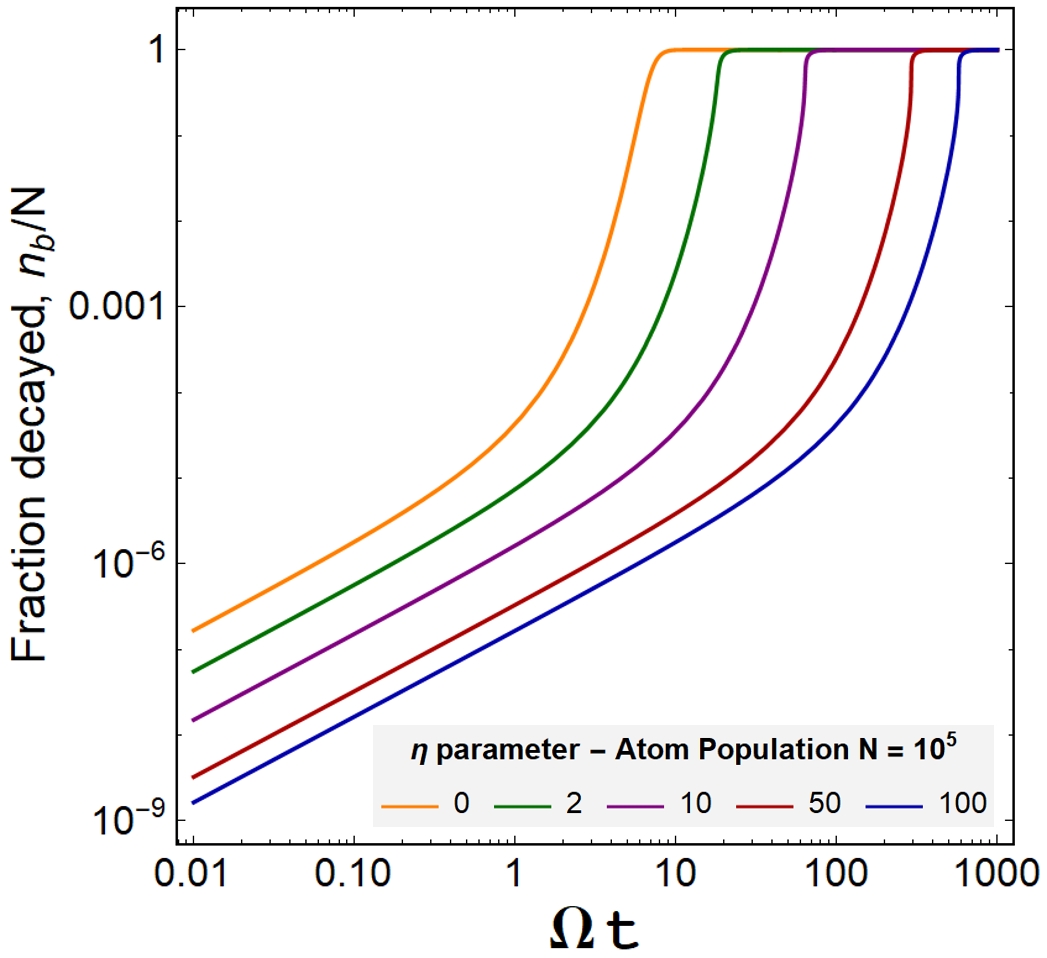}
    \caption{Plot of the decayed fraction $\frac{\braket{n_b}}{N}$, according to Eq.\eqref{eq:nbDynM}, for different values of the $\eta$ parameter for fixed atom population $N=10^5$.}
    \label{fig2}
\end{figure}

We now include short-range (contact) interactions between atoms through the Bose-Hubbard term 
\[
H_{\mathrm{coll}} = U_a n_a^2 + U_b n_b^2 + U_{ab} n_a n_b,
\]
which supplements the effective Hamiltonian presented in Eq. \eqref{eq:H_eff}. 
The contact atomic interaction, written in terms of the intra- and inter-species interaction strengths $U_{a,b}$ and $U_{ab}$, does not affect the populations dynamics, while provides modifications to the Heisenberg equation of motion of $S$ (see Appendix A). Indeed, the Heisenberg equation for the mixed correlator $S = \langle a^\dagger b c \rangle$ acquires an additional contribution from $H_{\mathrm{coll}}$, induced by
\begin{eqnarray}
[H_{\mathrm{coll}}, a^\dagger b c]
&=& (2U_a - U_{ab})\,a^\dagger b c\,n_a - (2U_b - U_{ab})\,a^\dagger b c\,n_b \nonumber\\
&+& (U_a + U_b-U_{ab})\,a^\dagger b c.
\end{eqnarray}
When all coupling constants are taken equal, $U_a = U_b = U_{ab} = U$, the result simplifies to
\[
[H_{\mathrm{coll}}, a^\dagger b c] = U\,a^\dagger b c\,(n_a - n_b + 1).
\]
Inserting this commutator into the Heisenberg equation for $S$ yields
\[
\dot S = \ldots + \frac{iU}{\hbar}\langle a^\dagger b c\,(n_a - n_b + 1)\rangle,
\]
where the dots denote the collisionless terms obtained previously. 
To close the hierarchy, we apply a mean-field decoupling,
\[
\langle a^\dagger b c\,(n_a - n_b+1)\rangle 
\simeq (\langle n_a \rangle - \langle n_b \rangle+1)\,S,
\]
neglecting higher-order correlations. 
With this approximation the equation for $S$ becomes
\[
\dot S = \frac{i}{\hbar}\Big[\Delta + U(\langle n_a \rangle - \langle n_b \rangle+1)+i \Gamma \Big]\,S + \ldots,
\]
so that the bare detuning $\Delta$ in Eq. (\ref{eq:SEOM}) is replaced by an effective, self-consistent quantity
\[
\Delta_{\mathrm{eff}} = \Delta + U(\langle n_a \rangle - \langle n_b \rangle+1).
\]
Hence, atomic collisions do not directly modify the population dynamics but shift the resonance condition through the instantaneous population imbalance between undecayed and decayed atoms. This interaction-induced shift acts as a nonlinear feedback on the cooperative emission, producing a self-consistent renormalization of the effective detuning during the condensate dynamics.\\
Proceeding under the same assumptions used in the absence of atom-atom interactions and considering the modifications induced by the atom-atom interaction on the dynamics of the mixed correlator $S$, we are able to derive a modified version of Eq. (\ref{eq:nbDyn}), which takes the following form:
\begin{eqnarray}
\label{eq:nbDynM}
 \braket{\dot{n}_b} =
     \frac{\Omega }{1+\eta \Big(1-2\frac{\langle n_b \rangle}{N} \Big)^2}  (1+\braket{n_b})(N-\braket{n_b}),
\end{eqnarray}
where we have introduced the dimensionless parameter $\eta=(UN/\Gamma)^2$. Close inspection of Eq. (\ref{eq:nbDynM}) shows that, in the relevant regime where $\eta \gg 1$, the decay dynamics is modified compared to the non-interacting case (i.e., $\eta=0$) in two ways.\\
At the process beginning, with small $\langle n_b \rangle$ values, the effective decay rate is lowered compared to the non-interacting case and tends to be $ \sim \Omega/\eta$, to be compared with the $\Omega$ value of the non-interacting case.\\
Moreover, the collective emission threshold, i.e. the threshold at which the decay process deviates from a pure exponential, shifts from $\langle n_b \rangle=N/2$ of the $\eta=0$ case to higher values depending on the strength of $\eta$. Interestingly, the collective emission threshold can be rigorously characterized by studying the equation $\langle \ddot{n}_b \rangle=0$ or by direct numerical simulation of Eq. (\ref{eq:nbDynM}).\\
We implemented the latter strategy and the results of this analysis, shown in Fig. \ref{fig2}, clearly evidence the impact of the atom-atom interaction on the system's dynamics.

\section{Neutrino emission in a Bosonic-Fermionic model}
\label{sec:BosFerm}
We now turn to the situation in which the initial atomic species is bosonic and forms a Bose-Einstein condensate, while the daughter atoms produced by electron capture are fermions. This case is of particular relevance, as emphasized in Ref.~\cite{Lu:2025sbp}, where it was shown that the fermionic statistics of the decay products leads to a severe suppression of superradiant amplification due to Pauli blocking.

In the following, we revisit this problem within the Lindblad formalism, which allows a consistent treatment of the irreversible character of the decay process and provides direct access to the dynamical mechanisms responsible for the inhibition of collective emission.

To this end, we describe the initial condensate by a bosonic annihilation operator $a$, while the fermionic daughter atoms populate an energy band labelled by single-particle modes $b_i$ with energies $E_i$. The emitted neutrino is represented by an operator $c$. Neglecting interactions other than those responsible for the decay, the free Hamiltonian reads
\begin{equation}
    H = E_a\, a^\dagger a 
      + \sum_i E_i\, b_i^\dagger b_i
      + E_\nu\, c^\dagger c ,
\end{equation}
where $E_a$ denotes the energy associated with the bosonic condensate mode, $E_i$ are the fermionic single-particle energies, and $E_\nu$ is the neutrino energy.

The decay process is described by the Lindblad jump operators:
\begin{eqnarray}
    L_\alpha^{(EC)} &=& \sqrt{g_\alpha} \, c^\dagger \, b_\alpha^\dagger \,a
    \\
    L_\nu^{(NE)} &=& \sqrt{\Gamma} \, c
    \\
    L_k^{(Th)} &=& \sqrt{\gamma_k}\,  b_{k-1}^\dagger \,b_k
\end{eqnarray}
where the index $\alpha$ denotes a preferred energy level within the energy band in which the atom decays.

A generic observable $O$ evolves according to the Lindblad-Heisenberg equation:
\begin{equation}
    \braket{\dot{O}} \,=\,\frac{i}{\hbar}\braket{[H,O]} \, +\, \braket{D(O)},
\end{equation}
where $D(O)=D(O)^{(EC)} +D(O)^{(NE)}+D(O)^{(Th)}$ denotes the Lindblad superoperator accounting for the electron capture process ($EC$), the neutrino emission ($NE$) and the thermalization ($Th$) of the decayed atoms :
\begin{eqnarray} \nonumber
    D(O)^{(EC)} &=& L_\alpha^{(EC)\dagger} \, O \,L_\alpha^{(EC)} \, - \, \frac{1}{2} \left\{
    L_\alpha^{(EC)\dagger}L_\alpha^{(EC)} , O\right\}
    \\ \nonumber
    D(O)^{(NE)} &=& L_\nu^{(NE)\dagger} \, O \,L_\nu^{(NE)} \, - \, \frac{1}{2} \left\{
    L_\nu^{(NE)\dagger}L_\nu^{(NE)} , O\right\}
    \\ \nonumber
   {}\hspace{-3mm} D(O)^{(Th)} \!&=&\!\sum_k\left[L_k^{(Th)\dagger} \, O \,L_k^{(Th)}  -  \frac{1}{2} \left\{
    L_k^{(Th)\dagger}L_k^{(Th)} , O\right\} \right].
\end{eqnarray}

Using the Lindblad-Heisenberg equation of motion the dynamics of the relevant operators $n_a=a^\dagger a$, $n_c=c^\dagger c$ and $n_k=b_k^\dagger b_k$ can be deduced. 

In particular, the equation for $n_a$ can be written as:
\begin{equation}
\label{eq:AtomDD}
    \braket{\dot{n}_a} \, = \, - g_\alpha \, \braket{a^\dagger a (1 -b_\alpha^\dagger b_\alpha ) (1 -c^\dagger c )},
\end{equation}

 while the equation for $n_c$ takes the form:
\begin{equation}\label{ncdot}
    \braket{\dot{n}_c} \, = \,  g_\alpha \, \braket{a^\dagger a (1 -b_\alpha^\dagger b_\alpha ) (1 -c^\dagger c )}\,-\,\Gamma \,\braket{n_c}.
\end{equation}

The thermalization process is described by the equations for $n_k$, which is written as:
\begin{eqnarray}\label{nkdot}
    \braket{\dot{n}_k} &= & \delta_{k\alpha} \,g_\alpha \, \braket{a^\dagger a (1 -b_\alpha^\dagger b_\alpha ) (1 -c^\dagger c )}
    \\ \nonumber
    &&-\gamma\braket{n_k(1-n_{k-1})} (1-\delta_{k,0})\,+\, \gamma\braket{n_{k+1}(1-n_k)},
\end{eqnarray}
where the label $k=0$ denotes the lowest single-particle fermionic level and we have assumed $\gamma_k \approx \gamma$.\\
The Eqs. (\ref{eq:AtomDD}), (\ref{ncdot}) and (\ref{nkdot}) do not form a closed set, as they couple to higher-order density-density correlations. In the present treatment, however, we neglect such correlations, thereby truncating the hierarchy and obtaining a self-consistent set of equations that can be solved numerically.\\
Considering the approximated form of the equations of motion, one can verify the conservation of the total (decayed and undecayed) number of atoms:
\begin{equation}
    \braket{\dot{n}_a} \, +\, \sum_k \braket{\dot{n}_k} \, = \,0.
\end{equation}

\begin{figure}[t]
    \includegraphics[width = 8.5 cm]{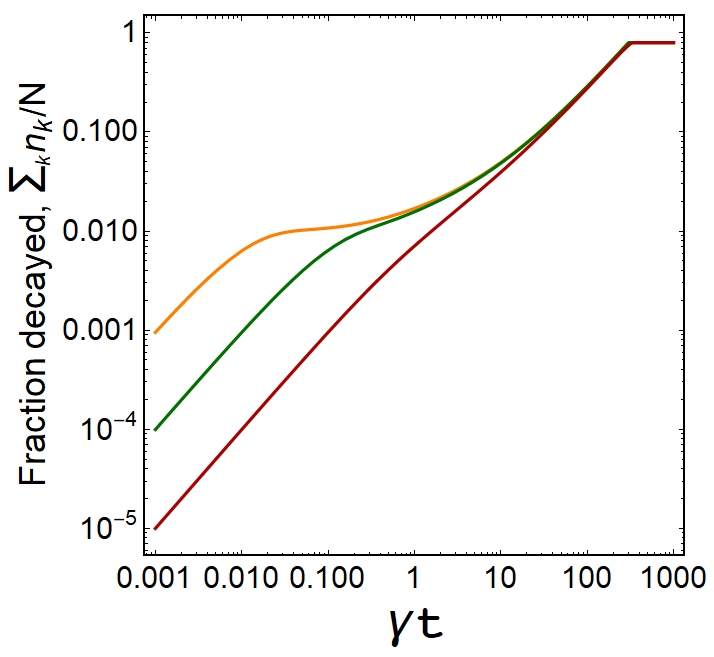}
    \caption{Plot of the decayed fraction $\sum_k\braket{n_k}/N$ as a function of time, obtained from  Eqs.\eqref{eq:AtomDD} and \eqref{nkdot} and assuming negligible density-density correlations ($\braket{n_k n_{k'}}\approx \braket{n_k}\braket{n_{k'}}$) and fast neutrino emission  ($\braket{n_c}\approx 0$). Distinct curves are obtained by fixing the atom population to $N=100$ and $\alpha=80$ and considering different values of the parameter $g_\alpha/\gamma$. We set $g_\alpha = \gamma$ for upper curve (yellow), 
    $g_\alpha = 0.1 \gamma$ for middle curve (green),
    $g_\alpha = 0.01 \gamma$ for lower curve (red).}
    \label{fig3}
\end{figure}

Moreover, close inspection of Eq. (\ref{eq:AtomDD}) reveals that, even under conditions of fast thermalization of the decayed atoms and rapid neutrino emission from the emitting region, the dynamics does not exhibit a superradiant behavior. Instead, it is well described by a simple exponential decay law, $\braket{\dot{n}_a} \approx -g_{\alpha} \braket{n_a}$. This result is further corroborated by numerical simulations of the decay dynamics, which confirm the detrimental influence of the Pauli blocking mechanism on the onset of collective emission effects.\\
The results of the latter numerical analysis are shown in Fig.~(\ref{fig3}), which illustrates the time evolution of the decayed fraction for different choices of the $g_{\alpha}/\gamma$ ratio. When thermalization process is slow, the occupation of the daughter fermionic level $\alpha$ builds up and produces an
intermediate-time Pauli-blocking plateau, temporarily suppressing further decay. This feature disappears when the relaxation dynamics is sufficiently fast to empty the preferential arrival level, thereby allowing decay to proceed
unimpeded. At long times, however, Pauli blocking inevitably arrests the dynamics once the fermionic levels become filled up to level $\alpha$.\\
In the specific cases shown in Fig.~(\ref{fig3}), this results in a residual population of about $20$ undecayed atoms out of $100$, explaining the asymptotic saturation observed in the long-time limit (i.e., $\gamma t \approx 10^3$).

These results agree with the general conclusions of  Ref.\cite{Lu:2025sbp}, even though we do not observe oscillating behaviour in the time evolution of the decaying population, which seems a consequence of the Hamiltonian approach there adopted.

\section{NEUTRINO EMISSION IN A
FERMIONIC BEC-BOSONIC BEC MODEL}
\label{sec:FermBos}
\begin{figure*}[t]
\includegraphics[width = 16 cm]{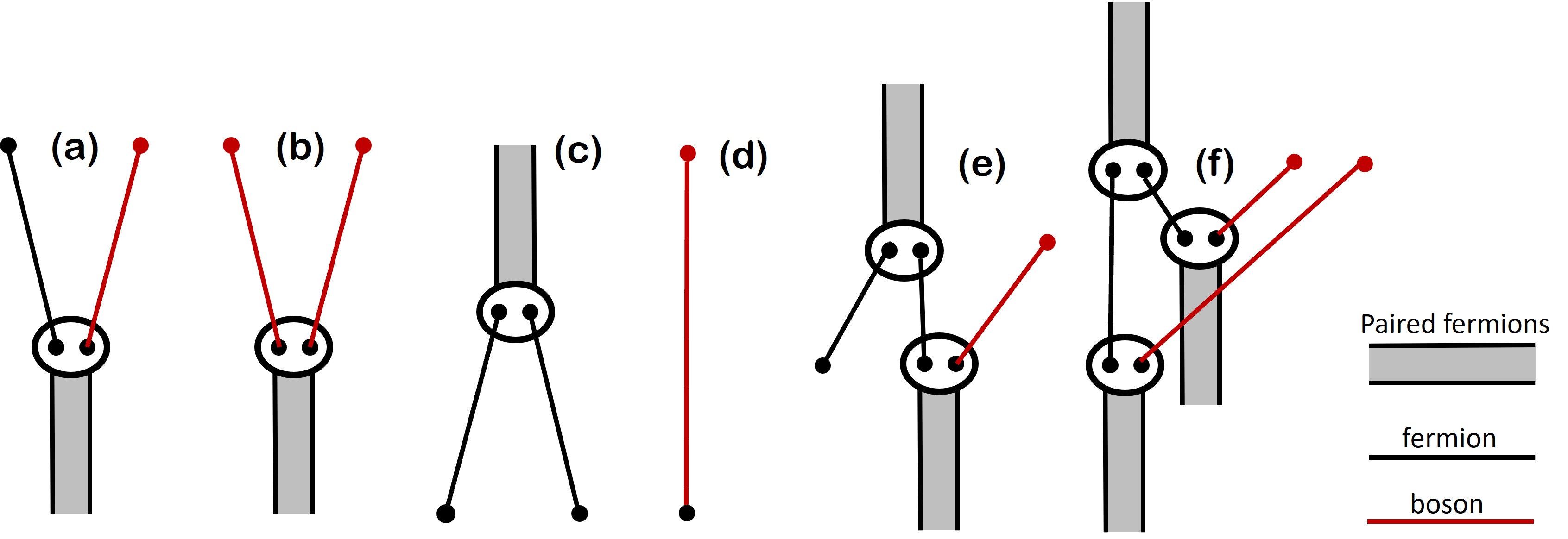}
\caption{Schematic representation of the elementary and composite processes occurring in a deeply bound fermionic BEC undergoing electron-capture decay into a bosonic atomic species. Double gray lines denote tightly bound fermion pairs, single black lines indicate unpaired fermions, and single red lines represent bosons produced by decay. Neutrino emission, always accompanying the decay process, is omitted for clarity. The physical regime considered assumes that the recombination time of two unpaired fermions is much shorter than the decay rate, so that the transient population of unpaired fermions is effectively negligible. (a) Decay of a single fermionic constituent within a bound pair, leading to the conversion of one fermion into a boson. (b) Double decay of both fermions forming a pair, resulting in the emission of two bosons. (c) Fast pairing (recombination) of two unpaired fermions into a tightly bound pair. (d) Decay of a single unpaired fermion into a bosonic atom. (e) Decay of one bound fermion within a pair followed by the recombination of the resulting unpaired fermion with another unpaired partner, forming a new pair. (f) A boson-mediated process in which the decay of a bound fermion and the subsequent recombination events effectively lead to the emission of two bosons, with the fermionic condensate providing the intermediate virtual channel.}
    \label{fig4}
\end{figure*}

Let us now consider an ultracold gas of fermionic atoms in the deep BEC regime, where tightly bound pairs behave as composite bosons and form a fermionic molecular condensate. In this regime, the condensation energy is typically much larger than any other dynamical scale associated with the decay process. As a consequence, any unpaired fermion generated by the decay of a molecule rapidly recombines with another unpaired atom, effectively suppressing the occupation of states with broken pairs. Two decay events occurring close in time would transiently produce two unpaired fermions, but these are energetically disfavored and promptly recondense (see Fig. \ref{fig4}). At the level of the effective low–energy theory projected onto the condensate subspace, the dynamical role of unpaired fermions is therefore negligible.

Under these conditions, the decay dynamics can be described in terms of an effective process in which a fermionic molecule is annihilated and two bosonic daughter atoms are created, with no long–lived intermediate states. The fermionic BEC then behaves as if the system explored virtual states with broken pairs only for a short time, before rapidly returning to the condensate manifold. This leads to a simplified description in which the dynamics is entirely encoded in a pair–destruction operator feeding two particles into the bosonic condensate.

For convenience, we introduce bosonic annihilation operators $a$ and $b$ describing, respectively, the fermionic-pair condensate and the daughter bosonic condensate, with single–mode energies $E_a$ and $E_b$. Although the operators associated with fermion pairs obey a hard–core boson algebra, the single–mode bosonic approximation provides an accurate description in the dilute deep–BEC regime, where deviations from canonical commutation relations are parametrically suppressed \cite{noziere}.

Within this coarse–grained framework, the irreversible dynamics of the decay can therefore be captured by an effective process in which the fermionic molecular mode $a$ is converted into two quanta of the bosonic mode $b$, consistently with the statistical and energetic structure of the system.

A minimal model for the description of such a process is the following:
\begin{eqnarray}
H &=& E_a \, a^\dagger a + E_b \, b^\dagger b \\ 
    L_D  &=& \sqrt{\gamma } \,  (b ^\dagger)^2 \,a
    \\
    L_d^{(a)} &=& \sqrt{\gamma_\phi^{(a)}} \,a^\dagger a
    \\
    L_d^{(b)} &=& \sqrt{\gamma_\phi^{(b)}}\,  b^\dagger \,b 
\end{eqnarray}
where the Hamiltonian part is diagonal in the two atomic species, while the irreversible part of the dynamics is modeled via jump operators, respectively, describing the decay ($L_D$) and  the decoherence in the fermionic BEC ($L_d^{(a)}$) and in the bosonic BEC  ($L_d^{(b)}$).

The degrees of freedom of the emitted neutrinos are neglected, assuming that they escape very rapidly from the emission region, thus ensuring that $\langle{n_c} \rangle\simeq 0$, as already seen in the previously discussed models.

The quantities of interest are therefore the atomic populations $\langle n_a \rangle$ and $\langle n_b \rangle$, with the constraint of the total atomic number conservation: 
\begin{equation} \label{number}
    2 \braket{n_a} +\braket{n_b} =N.
\end{equation}

The fact that $[H,n_a] = [H,n_b]=0$, implies that the dynamics of $n_a$ and $n_b$ is entirely regulated by the irreversible dynamics of the decay and by decoherence. 
In general, for an observable $O$, we have  LHEOM of the form:
\begin{equation}
    \braket{\dot{O}} \,=\,\frac{i}{\hbar}\braket{[H,O]} \, +\, \braket{D(O)},
\end{equation}
where $D(O)=D(O)^{(D)} +D(O)^{(d,a)}+D(O)^{(d,b)}$ with 
\begin{eqnarray}
 {}\hspace{-7mm}
 D(O)^{(D)}\!&=&   L_D^\dagger \, O\, L_D \, -\, \frac{1}{2}\Big\{L_D^\dagger \, L_D, \, O \Big\} 
 \\
 D(O)^{(d,\alpha)}\!&=&   L_D^{(\alpha)\dagger} \, O\, L^{(\alpha)}_D \, -\, \frac{1}{2}\Big\{L_D^{(\alpha)\dagger} \, L^{(\alpha)}_D, \, O \Big\} ,   
\end{eqnarray}
and $\alpha\in\{a,b\}$. Note that, if $O$ commutes both with $L$ and with $L^\dagger$, then $D(O)=0$.

Within the general framework evoked above, it is possible to derive the equation for $\braket{n_a}$ in the form:
\begin{eqnarray}
 \braket{\dot{n}_a} &=& \braket{D^{(D)}(n_a)} \, +\, \braket{D^{(d,a)}(n_a)} 
 \\ 
 &=& -\gamma \braket{n_a (n_b+1) (n_b+2)}.
\end{eqnarray}

On the other hand, the equation for $\braket{n_b}$ gives:
\begin{eqnarray}
 \braket{\dot{n}_b} &=& \braket{D^{(D)}(n_b)} \, +\braket{D^{(d,b)}(n_b)}
 \\ 
 &=& 2\gamma \braket{n_a (n_b+1) (n_b+2)},
\end{eqnarray}
therefore verifying the above stated conservation law, since $2\braket{\dot{n}_a}+\braket{\dot{n}_b}=0$.

Proceeding as done in Sec. \ref{sec:BosBos} and \ref{sec:BosFerm}, these equations can be solved by truncating correlations at this level, so that we obtain the following set of equations:
\begin{eqnarray}
 \braket{\dot{n}_a}
 &=& -\gamma \braket{n_a} (\braket{n_b}+1) (\braket{n_b}+2),
 \\ [2mm]
 \braket{\dot{n}_b} 
 &=& 2\gamma  \braket{n_a} (\braket{n_b}+1) (\braket{n_b}+2) ,
\end{eqnarray}
to be solved taking into account the total number conservation in Eq. \eqref{number} and the initial condition $\braket{n_b}|_{t=0}=0$.\\
Interestingly, within the truncation scheme adopted above, the jump operators $L_d^{(a,b)}$ related to the decoherence and proportional to $n_a$ and $n_b$, do not contribute to the closed dynamics of Eqs.~(40)–(41).\\
In a more refined treatment, however, decoherence would enter through the hierarchy of higher–order correlations. Indeed, although the time evolution of the correlator $\langle n_a (n_b+1)(n_b+2)\rangle$ is not directly modified by these dissipators, its equation of motion couples to additional mixed correlators whose dynamics would, in principle, be sensitive to the decoherence induced by the jump operators $L_d^{(a,b)}$.\\
These observations suggest that, although the decoherence channels would in principle affect the dynamics through such deeply nested correlators, their overall impact remains negligible at leading order, as already indicated by the truncated equations.

\begin{figure}[t]
    \includegraphics[width = 8.5 cm]{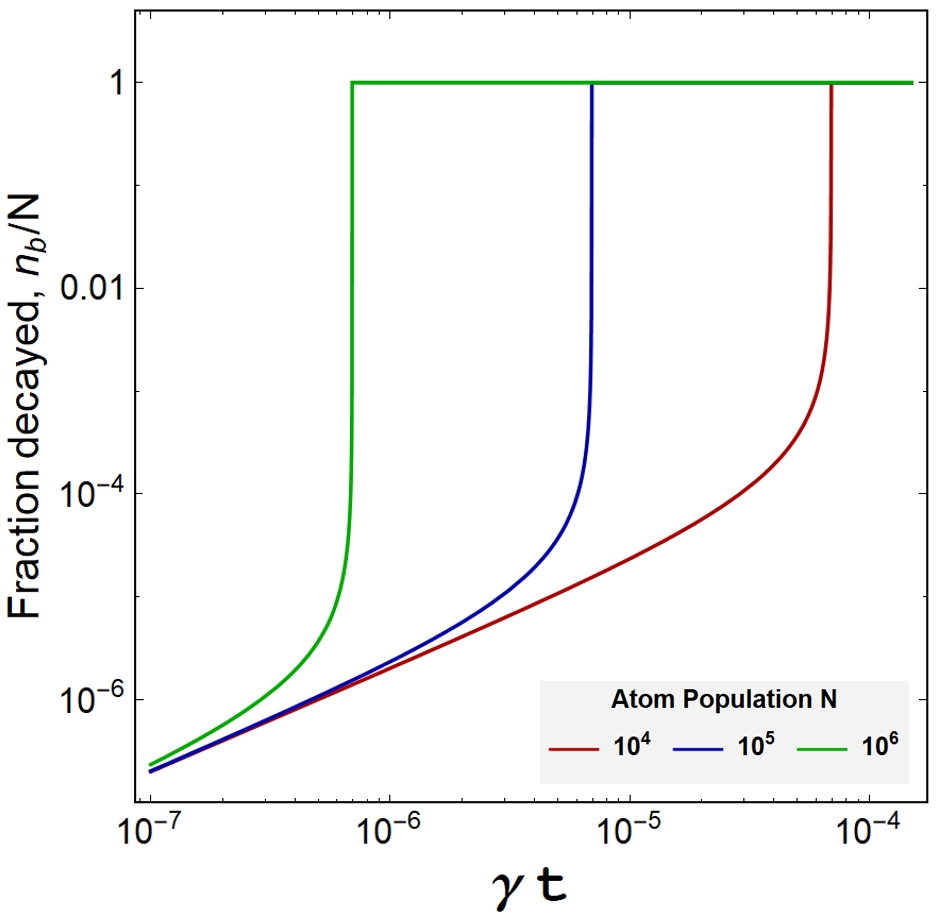}
    \caption{Plot of the decayed fraction $\frac{\braket{n_b}}{N}$, according to Eq.\eqref{nbFB}, for different atom populations $N$ as specified in figure legend.}
    \label{figFB}
\end{figure}

Thus, the resulting equation for $\braket{n_b}$ reads:
\begin{equation}
\label{nbFB}
    \braket{\dot{n}_b} 
 = \gamma  \left(N -\braket{n_b}\right)  (\braket{n_b}+1) (\braket{n_b}+2). 
\end{equation}
A direct comparison between the numerical integrations of Eq.~\eqref{nbFB} for
different atomic populations reveals the onset of a genuinely
\emph{collective} emission process, displaying a markedly
super-radiant character (see Fig. \ref{figFB}). Once $\langle n_b\rangle$ exceeds a threshold
value, the growth of the decayed fraction accelerates abruptly,
producing a dynamical ``explosion'' that is far more pronounced than
one would expect from a standard logistic-type equation, as typically
arises in mean-field models of boson-to-boson decay (see Section. \ref{sec:BosBos}). In the present
case, the combination of the bosonic stimulation factor
$(\langle n_b\rangle+2)$ in Eq.\eqref{nbFB} and the progressive
reduction of the available sites for decay leads to an autocatalytic
acceleration of the dynamics, culminating in a rapid collapse toward
saturation. The cooperative nature of the process manifests itself
through a sharp dynamical transition, which is sensitive to the total number of atoms. Indeed, increasing atomic
populations, the onset of the explosive growth shifts dramatically to
earlier times, while the initial evolution remains nearly identical
across all curves. This behaviour indicates that the collective
emission emerging in the model cannot be assimilated to a simple
logistic-like dynamics, but instead reflects the intrinsic bosonic
cooperativity embedded in the decay mechanism.

\section{Discussion and Conclusions}
\label{sec:Concl}
In this work, motivated by the recent proposal of Ref.~\cite{Jones:2024lhf}, we have undertaken a detailed analysis of the mechanism underlying superradiant neutrino emission from radioactive atomic condensates. Our first objective was to clarify the role of particle statistics in such a process.

We have shown that the scenario envisaged in Ref.~\cite{Jones:2024lhf} is, in principle, compatible with both the parent and daughter atoms being bosonic. However, this idealized situation is not realized in practice because the decay process leads to a change in statistics of the daughter species. In particular, starting from a BEC of bosonic atoms, the decay into fermionic atoms strongly suppresses the superradiant amplification, in agreement with the findings of Ref.~\cite{Lu:2025sbp}. Within the Lindblad formalism, we have examined this case in detail and confirmed the dominant role of Pauli blocking in quenching the cooperative emission, thereby reinforcing the conclusion that such a superradiant neutrino burst is effectively forbidden under these conditions.

We have then explored a complementary scenario in which the initial atomic population is fermionic and resides in the deep BEC regime, forming tightly bound pairs that decay into a bosonic condensate. Electron capture destabilizes the pairs, generating unpaired fermions; under the assumption of a rapid recombination of these atoms into new pairs, we have derived an effective model that does exhibit a pronounced superradiant behaviour, qualitatively consistent with the mechanism proposed in Ref.~\cite{Jones:2024lhf}. 

The analysis presented here is based on minimal models designed to isolate and highlight the statistical features that control the cooperative emission process. While simplified, these models capture the essential physics of the phenomenon and delineate the conditions under which superradiance may or may not occur. We expect that this study will help frame a more rigorous understanding of the underlying mechanisms, and more refined investigations are currently underway.

\appendix 
\section{Derivation of the effective model used in Sec. \ref{sec:BosBos}}
\label{app:realspace}
Hereafter, we provide details about the derivation of the effective model used in Sec. \ref{sec:BosBos} of the main text.\\
We consider a radioactive Bose-Einstein condensate undergoing electron capture, and construct an effective field-theoretic model retaining only the physically relevant degrees of freedom: two bosonic atomic fields $\psi_a(\mathbf r)$ and $\psi_b(\mathbf r)$ associated with the undecayed and decayed internal states, respectively, and a fermionic neutrino field $\nu(\mathbf r)$ describing the emitted neutrino modes. 

The total Hamiltonian of the emission region is thus written as
\begin{equation}
H = H_a + H_b + H_\nu + H_{\mathrm{int}} + H_{\mathrm{coll}}.
\end{equation}

The first two terms of $H$ describe the atomic kinetic and potential energies,
\[
H_\alpha = \int d^3r\,\psi_\alpha^\dagger(\mathbf r)\!\left[-\frac{\hbar^2\nabla^2}{2m_\alpha} + V_\alpha(\mathbf r)\right]\!\psi_\alpha(\mathbf r), \qquad \alpha = a,b,
\]
while the neutrino field contributes
\[
H_\nu = \int d^3r\, \nu^\dagger(\mathbf r)\,h_\nu\,\nu(\mathbf r),
\]
where $h_\nu$ denotes the single-particle Dirac Hamiltonian generating the free neutrino dispersion.  
Atom–atom interactions are included through the contact interaction term
\begin{eqnarray}
H_{\mathrm{coll}}&=&\tfrac{g_{aa}}{2}\int d^3r \ \psi_a^\dagger\psi_a^\dagger\psi_a\psi_a+\tfrac{g_{bb}}{2}\int d^3r \ \psi_b^\dagger\psi_b^\dagger\psi_b\psi_b+\nonumber \\
&+&g_{ab}\int d^3r \ \psi_a^\dagger\psi_b^\dagger\psi_b\psi_a. 
\end{eqnarray}
Atom decay and subsequent neutrino emission is described in effective manner tracing out the electronic degree of freedom, yielding an effective three-field vertex interaction of the form:

\begin{eqnarray}
H_{\mathrm{int}}=g_v\int d^3r\,[\,\mathcal \chi_{\nu}^\dagger(r)\,\psi_b^\dagger(r)\psi_a(r)+\mathrm{H.c.}\,],
\end{eqnarray}
where the quantity 
\begin{eqnarray}
\chi_{\nu}(r)=\sum_{\mathbf{k}s}\mathcal{F}_{\mathbf{k}s}(r)c_{\mathbf{k}s}
\end{eqnarray}
can be understood as a dressed field obtained as linear combination, with vertex functions $\mathcal{F}_{\mathbf{k}s}(r)$, of fermionic operators $c_{\mathbf{k}s}$ that destroy a positive energy solution of the single-particle Dirac equation $h_{\nu}u_{\mathbf{k}s}(r)=E_{\nu}(\mathbf{k})u_{\mathbf{k}s}(r)$. The same expansion is used to obtain the neutrino field as 
$\nu(\mathbf r)=\sum_{\mathbf k s}u_{\mathbf k s}(\mathbf r)\,c_{\mathbf k s}$ with fermionic commutation rule $\{c_{\mathbf k s},c_{\mathbf k' s'}^\dagger\}=\delta_{\mathbf k\mathbf k'}\delta_{ss'}$.\\
Projecting the atomic fields onto their condensate manifolds, we approximate each bosonic field by its lowest-energy single-particle mode, i.e. $\psi_\alpha(\mathbf r)\simeq\phi_\alpha(\mathbf r) \ \alpha$, with $\int d^3r \ |\phi_\alpha|^2=1$ and $\alpha\in \{a,b \}$. In the BEC regime this single-mode reduction is accurate, as population of higher orbital modes is energetically suppressed. Together with the scattering-eigenmode expansion of the neutrino field discussed above, the Hamiltonian reduces to the form
\begin{widetext}

\begin{eqnarray}
H= \varepsilon_a\,a^\dagger a+\varepsilon_b\,b^\dagger b+\sum_{\mathbf k,s}E_\nu(\mathbf k)\,c_{\mathbf k s}^\dagger c_{\mathbf k s}
+\sum_{\mathbf k,s}\!\Big[g_{\mathbf k s}\,c_{\mathbf k s}^\dagger\,b^\dagger a+g_{\mathbf k s}^*\,a^\dagger b\,c_{\mathbf k s}\Big]
+U_a\,(a^\dagger a)^2+U_b\,(b^\dagger b)^2+U_{ab}\,(a^\dagger a)(b^\dagger b),
\end{eqnarray}
\end{widetext}
where $\varepsilon_\alpha=\int d^3r\,\phi_\alpha^*(-\hbar^2\nabla^2/2m_\alpha+V_\alpha)\phi_\alpha$, $E_{\nu}(\mathbf{k})=\int d^3 r \ u^{\dagger}_{\mathbf{k}s}(r)h_{\nu}u_{\mathbf{k}s}(r)$,  $U_a=\tfrac{g_{aa}}{2}\int d^3r \ |\phi_a|^4$, $U_b=\tfrac{g_{bb}}{2}\int d^3r \ |\phi_b|^4$, $U_{ab}=g_{ab}\int d^3r \ |\phi_a|^2|\phi_b|^2$, and the microscopic vertex is
\begin{eqnarray}
g_{\mathbf k s}=g_{v}\int d^3r\; \mathcal{F}_{\mathbf k s}^\ast(\mathbf r)\,\phi_b^*(\mathbf r)\,\phi_a(\mathbf r).
\end{eqnarray}
 
Following Ref.~\cite{Jones:2024lhf}, we assume that the continuum of neutrino modes can be effectively projected onto a single dominant channel $(\mathbf{k}_0,s_0)$. Thus, the Hamiltonian in Eq. \eqref{eq:H_eff} is obtained by considering a single neutrino mode described by the second quantization operator $c\equiv c_{\mathbf{k}_0s_0}$ and energy $E_{\nu}=E_{\nu}(\mathbf{k}_0)$.\\
Using the Lindblad equation of motion for the relevant observables within the effective model framework and neglecting high-order correlations, the complete set of equations describing the system dynamics in the presence of atomic collisions, with $U_a=U_b=U_{ab}=U$, is obtained:
\begin{eqnarray}
\langle \dot{n}_a \rangle &=& \frac{2}{\hbar}\Im[g^\ast S]\\
\langle \dot{n}_b \rangle &=& -\frac{2}{\hbar}\Im[g^\ast S]\\
\langle \dot{n}_c \rangle &=& -\frac{2}{\hbar}\Im[g^\ast S]-\frac{2\Gamma}{\hbar}\langle n_c \rangle
\end{eqnarray}
\begin{widetext}
\begin{eqnarray}
\dot{S} &=& \frac{(i \Delta_{eff}- \Gamma)}{\hbar}S+\frac{ig}{\hbar}\Big[\langle n_c \rangle \langle n_b \rangle (1+\langle n_a \rangle)-(1- \langle n_c \rangle) \langle n_a \rangle (1+\langle n_b \rangle)\Big]
\end{eqnarray}
\end{widetext}
with $\Delta_{eff}=\Delta+U(\langle n_a \rangle-\langle n_b \rangle +1)$ and initial conditions $\langle n_a \rangle_{t=0}=N$ and $\langle n_b \rangle_{t=0}=\langle n_c \rangle_{t=0}=S(t=0)=0$.\\
The set of differential equations above can be integrated numerically in their full form, thus retaining the complete dimensionality of the problem. Alternatively, consistent reductions may be implemented by eliminating variables whose evolution occurs on faster time scales or remains only weakly coupled, as discussed in the main text. Both approaches yield mutually consistent evolutions, differing solely in the level of dynamical detail retained.

\end{document}